\documentclass{pasj00}
\draft

\begin{document}
\SetRunningHead{Author(s) in page-head}{Running Head}
%\Received{}%{yyyy/mm/dd}
%\Accepted{}%{yyyy/mm/dd}
%\Published{}%{yyyy/mm/dd}

\title{High-Resolution Near-Infrared Polarimetry of a Circumstellar Disk around UX Tau A}

%%% begin:list of authors
% Do NOT capitalize all letters in "textsc".
\author{
   Ryoko \textsc{Tanii}\altaffilmark{1},
   Yoichi \textsc{Itoh}\altaffilmark{1},
   Tomoyuki \textsc{Kudo}\altaffilmark{2},
   Tomonori \textsc{Hioki}\altaffilmark{1},
   Yumiko \textsc{Oasa}\altaffilmark{3},
   Ranjan \textsc{Gupta}\altaffilmark{4},
   A. K. \textsc{Sen}\altaffilmark{5},
% From here SEEDS co-authors
   J. P. \textsc{Wisniewski}\altaffilmark{6},
   T. \textsc{Muto}\altaffilmark{25},
   C. A. \textsc{Grady}\altaffilmark{9,10,11},
   J. \textsc{Hashimoto}\altaffilmark{12},
   M. \textsc{Fukagawa}\altaffilmark{13},
   S. \textsc{Mayama}\altaffilmark{14},
   J. \textsc{Hornbeck}\altaffilmark{15},
   M. \textsc{Sitko}\altaffilmark{16,17,18},
   R. \textsc{Russell}\altaffilmark{18,19},
   C. \textsc{Werren}\altaffilmark{17,18},
   M. \textsc{Cur\'e}\altaffilmark{20},
   T. \textsc{Currie}\altaffilmark{10},
   N. \textsc{Ohashi}\altaffilmark{2,21},
   Y. \textsc{Okamoto}\altaffilmark{22},
   M. \textsc{Momose}\altaffilmark{22},
   M. \textsc{Honda}\altaffilmark{23},
   S. -I. \textsc{Inutsuka}\altaffilmark{24},
   T. \textsc{Takeuchi}\altaffilmark{25},
   R. \textsc{Dong}\altaffilmark{26},
   L. \textsc{Abe}\altaffilmark{27},
   W. \textsc{Brandner}\altaffilmark{28},
   T. \textsc{Brandt}\altaffilmark{26},
   J. \textsc{Carson}\altaffilmark{29},
   S. \textsc{Egner}\altaffilmark{2},
   M. \textsc{Feldt}\altaffilmark{28},
   T. \textsc{Fukue}\altaffilmark{12},
   M. \textsc{Goto}\altaffilmark{28},
   O. \textsc{Guyon}\altaffilmark{2},
   Y. \textsc{Hayano}\altaffilmark{2},
   M. \textsc{Hayashi}\altaffilmark{30},
   S. S. \textsc{Hayashi}\altaffilmark{2},
   T. \textsc{Henning}\altaffilmark{28},
   K. W. \textsc{Hodapp}\altaffilmark{31},
   M. \textsc{Ishii}\altaffilmark{2},
   M. \textsc{Iye}\altaffilmark{12},
   M. \textsc{Janson}\altaffilmark{26},
   R. \textsc{Kandori}\altaffilmark{12},
   G. P. \textsc{Knapp}\altaffilmark{26},
   N. \textsc{Kusakabe}\altaffilmark{12},
   M. \textsc{Kuzuhara}\altaffilmark{12,30},
   T. \textsc{Matsuo}\altaffilmark{32},
   M. W. \textsc{McElwain}\altaffilmark{26},
   S. \textsc{Miyama}\altaffilmark{12},
   J. -I. \textsc{Morino}\altaffilmark{12},
   A. \textsc{Moro-Mart\'\i n}\altaffilmark{33},
   T. \textsc{Nishimura}\altaffilmark{2},
   T. -S. \textsc{Pyo}\altaffilmark{2},
   G. \textsc{Serabyn}\altaffilmark{8},
   H. \textsc{Suto}\altaffilmark{12},
   R. \textsc{Suzuki}\altaffilmark{12},
   M. \textsc{Takami}\altaffilmark{21},
   N. \textsc{Takato}\altaffilmark{2},
   H. \textsc{Terada}\altaffilmark{2},
   C. \textsc{Thalmann}\altaffilmark{28},
   D. \textsc{Tomono}\altaffilmark{2},
   E. L. \textsc{Turner}\altaffilmark{7,26},
   M. \textsc{Watanabe}\altaffilmark{34},
   T. \textsc{Yamada}\altaffilmark{35},
   H. \textsc{Takami}\altaffilmark{2},
   T. \textsc{Usuda}\altaffilmark{2},
   and M. \textsc{Tamura}\altaffilmark{12}
   }
 \altaffiltext{1}{Graduate School of Science, Kobe University, 1-1 Rokkodai-cho, Nada-ku, Kobe, Hyogo 657-8501}
 \email{tanii@stu.kobe-u.ac.jp}
 \altaffiltext{2}{Subaru Telescope, National Astronomical Observatory of Japan, 650 North A'ohoku Place, Hilo, HI 96720, USA}
 \altaffiltext{3}{Faculty of Education, Saitama University, 255 Shimookubo, Sakura, Saitama, Saitama 338-0825}
 \altaffiltext{4}{Inter University Center for Astronomy and Astrophysics(IUCAA), Ganeshkhind, Pune 411 007, India}
 \altaffiltext{5}{Department of Physics, Assam University, Silchar 788011, Assam, India}
 \altaffiltext{6}{Department of Astronomy, University of Washington, Box 351580 Seattle, WA 98195, USA}
 \altaffiltext{7}{Institute for the Physics and Mathematics of the Universe, The University of Tokyo, Kashiwa 227-8568, Japan}
 \altaffiltext{8}{Jet Propulsion Laboratory, California Institute of Technology, Pasadena, CA, USA}
 \altaffiltext{9}{Eureka Scientic, 2452 Delmer, Suite 100, Oakland CA 96002, USA}
 \altaffiltext{10}{ExoPlanets and Stellar Astrophysics Laboratory, Code 667, Goddard Space Flight Center, Greenbelt,MD 20771 USA}
 \altaffiltext{11}{Goddard Center for Astrobiology}
 \altaffiltext{12}{National Astronomical Observatory of Japan, 2-21-1 Osawa, Mitaka, Tokyo 181-8588, Japan}
 \altaffiltext{13}{Department of Earth and Space Science, Graduate School of Science, Osaka University, 1-1, Machikaneyama, Toyonaka, Osaka 560-0043, Japan}
 \altaffiltext{14}{The Graduate University for Advanced Studies (SOKENDAI), Shonan International Village, Hayama-cho, Miura-gun, Kanagawa, 240-0193, Japan}
 \altaffiltext{15}{Department of Physics and Astronomy, University of Louisville, Louisville, KY 40292, USA}
 \altaffiltext{16}{Space Science Institute, 4750 Walnut St., Suite 205, Boulder, CO 80301, USA}
 \altaffiltext{17}{Department of Physics, University of Cincinnati, Cincinnati, OH 45221-0011, USA}
 \altaffiltext{18}{Visiting Astronomer, NASA Infrared Telescope Facility, operated by the University of Hawaii under contract to NASA}
 \altaffiltext{19}{The Aerospace Corporation, Los Angeles, CA 90009, USA}
 \altaffiltext{20}{Departamento de F\'\i sica y Astronom\'\i a, Facultad de Ciencias, Universidad de Valpara\'\i so Av. Gran Breta\~na 1111, Casilla 5030, Valpara\'\i so, Chile }
 \altaffiltext{21}{Institute of Astronomy and Astrophysics, Academia Sinica, P.O. Box 23-141, Taipei 106, Taiwan}
 \altaffiltext{22}{Faculty of Science, Ibaraki University, 2-1-1 Bunkyo, Mito, Ibaraki, 310-8512, Japan}
 \altaffiltext{23}{Department of Information Science, Kanagawa University, 2946 Tsuchiya, Hiratsuka, Kanagawa 259-1293, Japan}
 \altaffiltext{24}{Department of Physics, Nagoya University, Furo-cho, Chikusa-ku, Nagoya, Aichi, 464-8602, Japan}
 \altaffiltext{25}{Tokyo Institute of Technology, 2-12-1 Ookayama, Meguro, Tokyo 152-8551, Japan}
 \altaffiltext{26}{Department of Astrophysical Sciences, Princeton University, NJ08544, USA}
 \altaffiltext{27}{Laboratoire Hippolyte Fizeau, UMR6525, Universite de Nice Sophia-Antipolis, 28, avenue Valrose, 06108 Nice Cedex 02, France}
 \altaffiltext{28}{Max Planck Institute for Astronomy, Heidelberg, Germany}
 \altaffiltext{29}{Department of Physics and Astronomy, College of Charleston, 58 Coming St., Charleston, SC 29424, USA}
 \altaffiltext{30}{Department of Astronomy, The University of Tokyo, Hongo 7-3-1, Bunkyo-ku, Tokyo 113-0033, Japan}
 \altaffiltext{31}{Institute for Astronomy, University of Hawaii, 640 North A'ohoku Place, Hilo, HI 96720, USA}
 \altaffiltext{32}{Department of Astronomy, Kyoto University, Kitashirakawa-Oiwake-cho, Sakyo-ku, Kyoto, 606-8502, Japan}
 \altaffiltext{33}{Department of Astrophysics, Center for Astrobiology, Ctra. de Ajalvir, km 4, Torrej\'on de Ardoz, 28850, Madrid, Spain}
 \altaffiltext{34}{Department of Cosmosciences, Hokkaido University, Sapporo, 060-0810, Japan}
 \altaffiltext{35}{Astronomical Institute, Tohoku University, Aoba, Sendai, 980-8578, Japan}

%\author{Ryoko \textsc{Tanii}\altaffilmark{1}, %
%  \thanks{Example: Present Address is xxxxxxxxxx}}
%\altaffiltext{1}{Graduate School of Science, Kobe University, 1-1 Rokkodai-cho, Nada-ku, Kobe, Hyogo 657-8501}
%\email{tanii@stu.kobe-u.ac.jp}
%
%\author{B-Firstname \textsc{B-Familyname}}
%\affil{B-Address of Institute}\email{bbbbb@xxx.xxx.xx.xx}
%\and
%\author{C-Firstname {\sc C-Familyname}}
%\affil{C-Address of Institute}\email{ccccc@xxx.xxx.xx.xx}
%%% end:list of authors

%%% Please use the following style in case that sorting by 
%%% affiliation is impossible. 
%
% \author{%
%   D-Firstname \textsc{D-Familyname}\altaffilmark{1}
%   E-Firstname \textsc{E-Familyname}\altaffilmark{1,2}
%   and
%   F-Firstname \textsc{F-Familyname}\altaffilmark{2}}
% \altaffiltext{1}{Address of Institute}
% \email{ddddd@xxx.xxx.xx.xx}
% \email{eeeee@xxx.xxx.xx.xx}
% \altaffiltext{2}{Address of Institute}

%% `\KeyWords{}' always has to be placed before `\maketitle'.
\KeyWords{(stars:) planetary systems: protoplanetary disk -- techniques: high angular resolution -- techniques: polarimetric} %Do NOT move this preamble from here!

\maketitle

\begin{abstract}
We present $H$-band polarimetric imagery of UX Tau A taken with HiCIAO/AO188 on the Subaru Telescope. 
UX Tau A has been classified as a pre-transitional disk object, 
with a gap structure separating its inner and outer disks. 
Our imagery taken with the 0.15$\arcsec$ (21 AU) radius coronagraphic mask has revealed a strongly polarized circumstellar disk surrounding UX Tau A 
which extends to 120 AU, at a spatial resolution of 0.1$\arcsec$ (14 AU).
It is inclined by 46$\degree$ $\pm$ 2$\degree$ as the west side is nearest. 
Although SED modeling and sub-millimeter imagery suggested the presence of a gap in the disk, 
with the inner edge of the outer disk estimated to be located at 25 -- 30 AU, 
we detect no evidence of a gap at the limit of our inner working angle (23AU) at the near-infrared wavelength. 
We attribute the observed strong polarization (up to 66 \%) to light scattering by dust grains in the disk.
However, neither polarization models of the circumstellar disk based on 
Rayleigh scattering nor Mie scattering approximations were consistent with the observed azimuthal profile of the polarization degrees of the disk. 
Instead, a geometric optics model of the disk with nonspherical grains with the radii of 30 $\mu$m is consistent with the observed profile. 
We suggest that the dust grains have experienced frequent collisional coagulations and have grown in the circumstellar disk of UX Tau A.

\end{abstract}

\section{Introduction}
%% planetary formation process
A protoplanetary disk around a young stellar object is the site of planetary formation. 
The core accretion model (e.g., \cite{Nakagawa83}) and the gravitational instability model (e.g., \cite{Boss98}) 
are two possible methods by which the process of planet formation might occur. 
In the core accretion model, a key process is dust grain growth by collisional coagulations.
Several fundamental processes of the dust growth have been proposed under simple assumptions.
Dust grains with sizes less than a few tens of micrometers
orbit a central star with the same velocity as gas (\cite{Adachi76}). 
They frequently collide and coagulate with each other. Simultaneously they begin to settle 
into an equatorial plane of the disk (\cite{Nakagawa81}). Dust grains with different settling velocities 
also experience collisional coagulations. As a result, planetesimals (radii $\sim$ 10 -- 100 km; \cite{Kokubo98}) are formed in the mid-plane of the disk. 
Bodies between a few tens of micrometers and a few kilometers are dominated by gas drag.
Such bodies are thought to migrate towards the central star in a short timescale (e.g., \cite{Nakagawa86}).
This rapid inward migration poses a challenge to planet formation theory.
By contrast, for planetesimals larger than a few kilometers, 
gravitational interactions are significant and relative velocities between the bodies increase due to mutual perturbations. 
This situation allows faster growth of the larger bodies and leads to formation of planetary embryos (radii $\sim$ 1000 km; \cite{Kokubo00}). 
The core accretion model is one of the plausible model of planetary formation. 
However it is still debated how sub-micron sized grains are transformed into kilometer sized bodies. 

%% transitional disk objects
An infrared excess in the spectral energy distribution of a young stellar object provides indirect evidence for the presence of a circumstellar disk. 
The spectral energy distributions of T Tauri stars with continuous, optically thick disks have strong infrared excesses 
in the near-infrared to far-infrared wavelengths (\cite{Williams11}). 
Recently, objects with large continuum excesses in the mid- to far-infrared wavelengths 
but no excess in the near-infrared wavelengths have been discovered, called 'transitional disk objects' (TDOs; \cite{Calvet02}, \cite{Calvet05}). 
Such an object is considered to lie in the transition state 
between classical T Tauri stars (CTTSs) and weak-line T Tauri stars (WTTSs). 
It is expected that the inside of the disk has been cleared out by dust accumulations and/or formations of protoplanets. 
Photo-evaporation process is also proposed as a mechanism to create a transitional disk. 
On the other hand, 'pre-transitional disk objects' (PTDOs) are considered to be in the evolutionary phase 
before reaching the transitional disk phase. They have mid- to far-infrared excesses similar to TDOs, 
however, they also show small excess in the near-infrared wavelengths (\cite{Esp07}). 
The slight near-infrared excess implies that optically thick material remains in the innermost part of the disk. 
\citet{Esp10} proposed that the planetary formation process is expected to progress in the gap structures between the inner and outer disks. 

High spatial resolution coronagraphic imaging polarimetry is one way to directly diagnose young circumstellar disks. 
The degree of polarization depends on scattering angle, grain size, and composition. 
By constructing the polarization profile of the disk, we are able to investigate size and composition of dust grains. 
\citet{Silber00} conducted infrared polarization imaging observations of a circumbinary disk around GG Tau 
with NICMOS mounted on the Hubble Space Telescope. 
The circumbinary disk shows strong polarization degrees of $\sim$ 50\% at 1 $\mu$m wavelength. 
The polarization azimuthal profile indicated Rayleigh-like scattering from dust grains with sub-micron size. 
\citet{Hashimoto11} conducted $H$-band polarization imaging observations of AB Aur with HiCIAO/AO188 on the Subaru Telescope
and revealed spiral structure in the outer part and the double ring structure at the inner part of the circumstellar disk. 
There was a number of studies in the last years, using the same observational technique for similar sources, 
e.g., \authorcite{Apai04} (\yearcite{Apai04}; TW Hya), \authorcite{Oppenheimer08} (\yearcite{Oppenheimer08}; AB Aur), 
\authorcite{Perrin09} (\yearcite{Perrin09}; AB Aur), \authorcite{Quanz11} (\yearcite{Quanz11}; HD100546), \authorcite{Quanz12} (\yearcite{Quanz12}; HD97048). 

%% UX Tau basical information
UX Tau is a T Tauri multiple system (\cite{Jones79}) in the Taurus molecular cloud (distance $\sim$ 140 pc; \cite{Elias78}). 
It consists of a primary star (UX Tau A), with UX Tau B separated by 5.86$\arcsec$ and UX Tau C separated by 2.63$\arcsec$ from the primary. 
UX Tau B is itself a binary system with the separation of 0.14$\arcsec$. The spectral type of UX Tau A is K2 (\cite{Kraus09}). 
Its spectral energy distribution shows a slight excess in the near-infrared wavelengths and 
significant excesses in the mid- and far-infrared wavelengths. These characteristics indicate 
that UX Tau A has an optically thick inner disk separated from an optically thick outer disk by a gap, 
i.e. it is a pre-transitional disk object (\cite{Esp10}). 
Model fits of the SED suggested that the outer edge of the inner disk is located 
at $<$ 0.21 AU and the inner wall of the outer disk is located at 30 AU from the central star (\authorcite{Esp10} \yearcite{Esp10}, \yearcite{Esp11}). 
The disk around UX Tau A was spatially resolved by Sub-millimeter Array at 880 $\mu$m wavelength, 
with the spatial resolution of 0.3$\arcsec$ (\cite{And11}). They found a dust-depleted disk cavity around the central star, 
and estimated the inner edge of the outer disk to be located 25 AU from the central star. 
UX Tau A shows no 10 $\mu$m silicate emission (\cite{Esp10}), implying a lack of small dust grains in the disk. 
These features suggest dust grain growth in the circumstellar disk of UX Tau A.
We conducted polarization imaging observations of UX Tau A in the $H$-band (1.6 $\mu$m) 
and investigate the collisional coalescence process of the dust grains in its disk.

%\noindent IMPORTANT NOTICE\\
%1. ``\verb|\draft|'' creates single column and double spaces format.\\
%2. If you comment out ``\verb|\draft|'', the output will be double column
%   and single space.\\
%3. For cross-references, the use of ``\verb|\label|, \verb|\ref|, \verb|\cite|'' 
%   and the thebibliography environment is strongly recommended. \\
%4. Do NOT use ``\verb|\def|, \verb|\renewcommand|''.\\
%5. Do NOT redefine commands provided by PASJ00.cls.\\

\section{Observations}

Near-infrared ($H$-band; 1.6 $\mu$m) polarimetric imaging observations of UX Tau A were carried out 
2009 December 23 with HiCIAO (High Contrast Instrument for the Subaru next generation Adaptive 
Optics; \cite{Tamura06}) and the adaptive optics system, AO188 (\cite{Hayano10}), 
mounted on the Nasmyth platform of the Subaru Telescope (Table 1). 
These observations were conducted as part of the larger SEEDS survey (\cite{Tamura09}). 
We employed the Polarization Differential Imaging (PDI) mode. 
The Wollaston prism installed in HiCIAO divides incident light 
into two linearly polarized components which are perpendicular to each other 
and which are imaged simultaneously on the detector. 
Each image has 1024 $\times$ 2048 pixels with a field of 
view of 9.75$\arcsec$ $\times$ 20.09$\arcsec$ and pixel scale of 9.521 mas/pixel in the east-west direction and 9.811 mas/pixel 
in the north-south direction. In the PDI mode, when the half-wave plate is set at the offset 
angle of 0$\degree$, 45$\degree$, 22.5$\degree$, and 67.5$\degree$, 
we obtain polarimetric images with the polarization direction at 
0$\degree$ and 90$\degree$, 90$\degree$ and 0$\degree$, 45$\degree$ and 135$\degree$, and 135$\degree$ and 45$\degree$ components respectively.
We used a coronagraphic mask with 0.3$\arcsec$ diameter to suppress the brightness of UX Tau A. 
We obtained 44 frames, i.e. 11 frames per half-wave plate position,
of UX Tau A with an exposure time of 60 s. 
The outside of the coronagraphic mask was not saturated. 
During the observations we fixed the star at the center of the coronagraphic mask, i.e. dithering method was not employed. 
In these frames, the tertiary component (UX Tau C) was also imaged outside the coronagraphic mask. 
The PSF reference star SAO93770 was imaged before observing UX Tau, using the PDI mode with the 0.3$\arcsec$ diameter coronagraphic mask. 
The natural seeing was 0.5 -- 0.6$\arcsec$ in the $K$-band. 
The AO188 measures the wavefront distortion by the atmospheric turbulence in the $R$-band wavelength 
and compensates it at all wavelength. 
The FWHM of UX Tau C was 10 pixels ($\sim$ 0.1$\arcsec$) on average. 
In the observations of SAO93770, 
with the ND filter in the AO device, we set the brightness difference between it and UX Tau to be 0.07 mag in the $R$-band 
in order to level the AO correction performance in the observations of the object and the reference star. 
We took 11 frames with an exposure time of 90 s without employing dithering method. 
We obtained 10 flat frames on the same day as the target observations 
and 49 dark frames in 2009 December 24. Each exposure time was 80 s for the flat frames and 5.6 s for the dark frames. 
The dark frames were used only for the identification of hot pixels. 

\begin{table}
  \caption{Properties of the observing objects}\label{tab:table01}
  \begin{center}
    \begin{tabular}{llllll}
      \hline \hline
      Target & RA [J2000] & DEC [J2000] & Sp.Type\footnotemark[$*$] & H [mag]\footnotemark[$*$$*$] & Airmass\\ 
      \hline
      UX Tau A & 04 30 03.991 & +18 13 49.39 & K2 & 8.0 & 1.2 -- 1.6\\
      UX Tau C & 04 30 03.991 & +18 13 49.39 & M5 & 10.9 & 1.2 -- 1.6\\
      SAO93770 & 04 07 08.734 & +10 47 58.86 & F8 & 7.0 & 1.2 -- 1.3\\
      \hline 
      \multicolumn{6}{@{}l@{}}{\hbox to 0pt{\parbox{120mm}{\footnotesize
        \par\noindent
        \footnotemark[$*$] \citet{Kraus09} for UX Tau A and C. The Hipparcos and Tycho Catalogues for SAO93770.
        \footnotemark[$*$$*$] \citet{Correia06} for UX Tau A and C. NOMAD catalog for SAO93770.
        }\hss}}
    \end{tabular}
  \end{center}
\end{table}

\section{Data Reduction}

The object frames were calibrated with the Image Reduction and Analysis Facility (IRAF). 
All frames of HiCIAO/AO188 have artifacts of horizontal stripes and vertical bandings. The horizontal stripe 
has a size of 2048 $\times$ 64 pixels. Each frame has 32 horizontal stripes. 
We attribute the origin of these artifacts to a bias instability caused by temperature fluctuations between the detector and the 
Application Specific Integrated Circuit (ASIC). The ASIC translates the detector's analog signals 
into the digital signals. The vertical bandings occur at one pixel interval. They show an alternate 
direction between even and odd horizontal stripes. 

To mitigate these artifacts, first, the horizontal stripes were removed from the dark frames. 
We measured a median count for each horizontal stripe and subtracted it from each horizontal stripe. 
For the vertical bandings, even horizontal stripes were vertically flipped. We median-combined 32 horizontal stripes 
to make a master horizontal stripe. We subtracted the master stripe from each odd horizontal stripe. 
From each even horizontal stripe, the vertically flipped master stripe was subtracted. 
Finally, the dark frames were median-combined. 
This combined dark frame was only used as a mask frame for the extraction of hot pixels.

Next, we eliminated hot and bad pixels from the object and flat frames. 
We regarded the pixels with more than 100 ADU in the combined dark frame as hot pixels. 
It is nearly equivalent to  12$\sigma$ above the median of the combined dark frame. 
Hot pixels were replaced with 10000 ADU and the other pixels were replaced with 0 ADU. 
Using this frame as a hot pixel mask, hot pixels were interpolated by nearest good pixels in the object and flat frames. 
Then, we removed the stripe patterns from the object frames with the same procedure we used for the dark frames. 
In this process, however, we put circular masks on the stars and computed the median values of the stripes and the bandings except the mask regions. 
We considered that dark counts were also subtracted from the object frames through this destriping process. 
We did not subtract the stripes from the flat frames. Because the flat frames had 
counts two orders of magnitude larger than the stripes, we ignored the stripe modulations. 
The flat frames were median-combined and normalized to 1 ADU. 
The object frames were divided by the flat frame. Bad pixels and cosmic rays were interpolated by nearest good pixels in the object frames. 

After those processes, we obtained the flux images of the polarimetric components, $F_{0\degree}$ and $F_{90\degree}$, $F_{90\degree}$ and $F_{0\degree}$, 
$F_{45\degree}$ and $F_{135\degree}$, and $F_{135\degree}$ and $F_{45\degree}$, 
separated into left and right of the image by $\sim$ 1070 pixels with each other. 
We cut the object frames into the left and right images. 
Each image has a 9.75$\arcsec$ $\times$ 20.09$\arcsec$ field of view. 
For each image, distortions caused by the Wollaston prism was corrected. 
Then, the central coordinates of the companion, UX Tau C, in all object images were measured. 
Using those coordinates, the object images were aligned. We also carried out the same procedures 
for the PSF reference star frames. 
The position of the photo-center of the reference star was estimated by fitting the halo component with a 2D Gaussian profile. 
Two frames of UX Tau A and three frames 
of the PSF reference star were not included in the following process, since they had insufficient tip-tilt correction. 

Linear polarization is described by the degree of polarization $P$ and polarization angle $\alpha$. 
These values are derived with the Stokes parameters, $I$, $Q$, and $U$\\
\begin{equation}
	\frac{Q}{I}=\frac{F_{0\degree}-F_{90\degree}}{F_{0\degree}+F_{90\degree}} \label{eq:eqQ}
\end{equation}
\begin{equation}
	\frac{U}{I}=\frac{F_{45\degree}-F_{135\degree}}{F_{45\degree}+F_{135\degree}}\label{eq:eqU}
\end{equation}
\begin{equation}
	P=\sqrt{\left(\frac{Q}{I}\right)^{2}+\left(\frac{U}{I}\right)^{2}}\label{eq:eqP}
\end{equation}
\begin{equation}
	\alpha=\frac{1}{2}\arctan{\left(\frac{U}{Q}\right)}\label{eq:eqAlpha},
\end{equation}
where $F_{0\degree}$, $F_{90\degree}$, $F_{45\degree}$, and $F_{135\degree}$ are the intensity of each polarimetric component.
By subtracting the right images from the left images, we derived 11 images of $Q$ and $-Q$, and 10 images of $U$ and $-U$. 
This subtraction method suppressed speckle noises that appeared both in the right and left images. 
We subtracted the $-Q$ images from the $Q$ images and the $-U$ from the $U$. 
Thus, different aberrations between the left and right images 
due to the different dividing directions of the Wollaston prism were cancelled (double-difference technique; \cite{Hinkley09}). 
2$Q$ and 2$U$ images were divided by 2, then the $Q$ and $U$ images were combined respectively. 
Then, we corrected the instrumental polarization
induced from the telescope and the instruments.
We refer to Joos et al. (2008) for the correction.
First, we construct a 2 $\times$ 2 Jones matrix
for each optical element.
In the HiCIAO case, we consider that
the tertiary mirror and three mirrors of the derotator
are probably dominant sources of the instrumental polarization.
The surface materials of the mirrors expressed by the complex
reflection index ($n = \eta$ + $i\kappa$) was
actually measured by the ellipsometer.
We also describe the rotation of Stokes vectors as a function of
the parallactic angle. Then, the multiplication of
all single Jones matrices leads to a Jones matrix of
whole telescope/instrument for a given set-up.
Finally, the constructed Jones matrix was
inverted and multiplied with the measured Stokes vectors
to obtain the original Stokes vectors on the sky.
In above method, we assumed monochromatic light and
infinite F-number.

These images had distortions by the compensator in HiCIAO, the AO system, and the telescope. 
The compensator is located in front of the focal mask and reduces chromatic aberrations. 
By using images of M15 observed by the HST and HiCIAO, we calibrated the image distortion 
of $Q$ and $U$. Multiplying both sides of equation (\ref{eq:eqP}) by $I$, we obtain the following equation, 
\begin{equation}
	PI=\sqrt{Q^{2}+U^{2}}\label{eq:eqPI} .
\end{equation}
We constructed polarized intensity ($PI$) images with the $Q$ and $U$ images. Assuming that polarization 
due to the interstellar materials in front of UX Tau is negligible (See Section 4), 
the polarized intensity image represents the polarized component of circumstellar structures around the central star.

In order to derive the polarization $P$ of the circumstellar structures, the $PI$ image needs 
to be divided by the $I$ image which contains only the components of the circumstellar structures. 
The $I$ image of UX Tau A (hereafter $I_{\rm tot}$) consists of the intensity of the circumstellar structures as well as that of the central star. 
By subtracting the $I$ component of the central star (hereafter $I_{\rm *}$) from the $I_{\rm tot}$, 
we obtain the $I$ image of the circumstellar structures (hereafter $I_{\rm disk}$), i.e. $I_{\rm disk} = I_{\rm tot} - I_{\rm *}$. 
We adopted the $I$ image of the PSF reference star (hereafter $I_{\rm psf}$) as the $I_{\rm *}$. 
First, we constructed $I$ images. We used the left and right images, 
$F_{0\degree}$ and $F_{90\degree}$, $F_{90\degree}$ and $F_{0\degree}$, $F_{45\degree}$ and $F_{135\degree}$, and $F_{135\degree}$ and $F_{45\degree}$. 
For UX Tau, we combined the left and right images into $I_{\rm tot}$ images. 
For the PSF reference star, the central star position was measured in each image with fitting the halo component with a 2D Gaussian profile, 
then the position offset was corrected. 
Then we combined the left and right images into the $I_{\rm psf}$ image. 
The image distortion was also calibrated. Two frames of the PSF reference star have distorted PSFs, 
so that we used six frames of the PSF reference star in the following procedure.

Since the Subaru Telescope is an alt-azimuth telescope and HiCIAO is mounted on the Nasmyth platform, 
the position angle of the spider changes in each image. 
The $I_{\rm psf}$ images were rotated to adjust their spider directions to those of every $I_{\rm tot}$ image. 
The six $I_{\rm psf}$ images of the PSF reference star were median-combined to make a PSF template $I$ image (hereafter $I_{\rm *tmp}$) for each $I_{\rm tot}$ image. 
Finally, we made 42 $I_{\rm *tmp}$ images. 
From these $I_{\rm tot}$ and $I_{\rm *tmp}$ images, sky levels were subtracted. 
The sky value was the average of the mean counts of four regions. 
The regions were selected as the 100 $\times$ 100 pixel regions separated at least 5$\arcsec$ from the central star. 
Next, we made the $I_{\rm disk}$ image. In each $I_{\rm tot}$ and $I_{\rm *tmp}$ image, 
we measured halo intensities of the central star. The evaluated regions were four regions of 20 $\times$ 20 pixels, which had a distance of 150 pixel (1.47$\arcsec$) 
from the central star and did not overlap the spider directions. 
We used the ratio of the mean values of these halos to determine the proper scaling to apply to $I_{\rm *tmp}$ image during PSF subtraction. 
We also shifted $I_{\rm *tmp}$ image slightly so that intensity of the circumstellar disk would not show strong asymmetry in the residual image. 
We assume that the disk intensity is symmetric, thus we may have over-corrected the PSF subtraction. 
The amount of the shift was 1.3 pixels (0.01$\arcsec$ = 0.1 FWHM $\sim$ 2 AU) on average. 
Therefore, the inner working angle of the $I_{\rm disk}$ image is $\sim$ 23 AU. 
Then, we subtracted $I_{\rm *tmp}$ image from $I_{\rm tot}$ image. This process was performed in each $I_{\rm tot}$ image. 
As a result, we obtained 42 $I_{\rm disk}$ images containing only the circumstellar structure. They were median-combined. 
Finally, the $PI$ image was divided by the $I_{\rm disk}$ image, thus the $P$ image of the circumstellar structure was generated.

\section{Results}

\begin{figure}
  \begin{center}
    \FigureFile(80mm,80mm){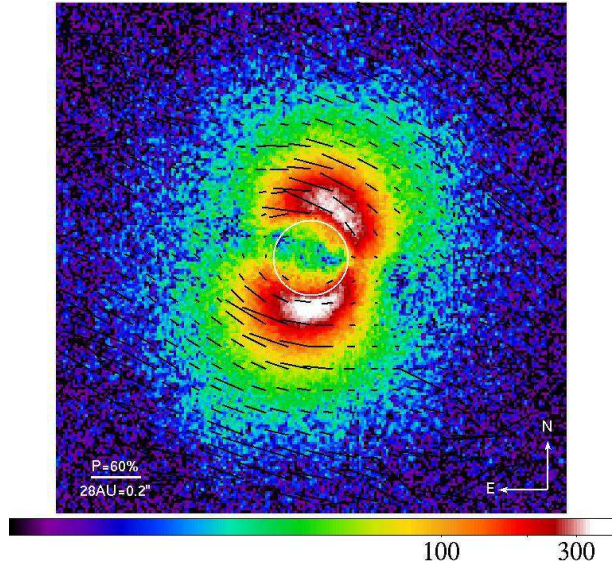}
  \end{center}
  \caption{
  	A polarized intensity image of UX Tau A. 
        The polarization vectors are superimposed in the region where the polarized intensity is larger than 5$\sigma$. 
        The vector lengths and directions indicate the degree and angle of the polarization, respectively. 
        The vector scale is shown in the bottom-left. A circle shows the 0.3$\arcsec$ diameter coronagraphic mask positioned on the central star. 
        The bottom color-bar denotes the polarized intensity in ADU. 
        }\label{fig:PIimg}
\end{figure}

%% main comments about PI image
We detected strongly polarized components around UX Tau A. The polarized intensity ($PI$) image 
shows an elliptical structure (Figure \ref{fig:PIimg}). 
We evaluate the extent of the region where the intensity is more than five times the standard deviation in the sky region of the $PI$ image. 
It has a semi-major axis of $\sim $0.83$\arcsec$ ($\sim$ 120AU) with the position angle of 165$\degree$ $\pm$ 2$\degree$ 
and a semi-minor axis of 0.58$\arcsec$ ($\sim$ 80AU). 

\begin{figure}
  \begin{center}
    \FigureFile(70mm,70mm){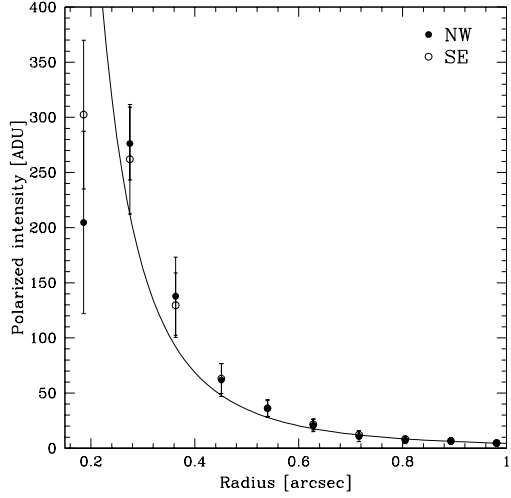}
  \end{center}
  \caption{
  	Radial profiles of the polarized intensity measured along semi-major axes of the circumstellar disk of UX Tau A. 
        The plots and the errorbars show mean polarized intensity and standard deviation in every 9 $\times$ 9 pixel region. 
        The filled and open circles denote the polarization intensity along semi-major axes in the north-west and south-east directions, respectively.
        The solid line show a power law fitting to the observed radial profile. The polarized intensity declines as the third power of the disk radius. 
        }\label{fig:PIradprof}
\end{figure}

We associate this polarization structure to the circumstellar disk of UX Tau A. 
If the circumstellar disk of UX Tau A has a circular and geometrically thin structure, 
the disk is tilted to east-west with the inclination of 46$\degree$ $\pm$ 2$\degree$.
The west side of the structure is brighter than the east side. 
We assume that the west side of the structure is the near side to us.
We measured degrees and angles of polarization in every 9 $\times$ 9 pixel region. 
The degree of polarization in regions with SNR $>$ 5$\sigma$ ranges from 1.6 to 66 \%. 
The polarization vectors show a mostly centrosymmetric pattern centered on UX Tau A, 
supporting our claim that this structure is attributable to its circumstellar disk. 
The radial profiles of the polarized intensity along the semi-major axes are shown in Figure \ref{fig:PIradprof}. 
The polarized intensity declines with the third power of the disk radius. 

\begin{figure}
  \begin{center}
    \FigureFile(70mm,70mm){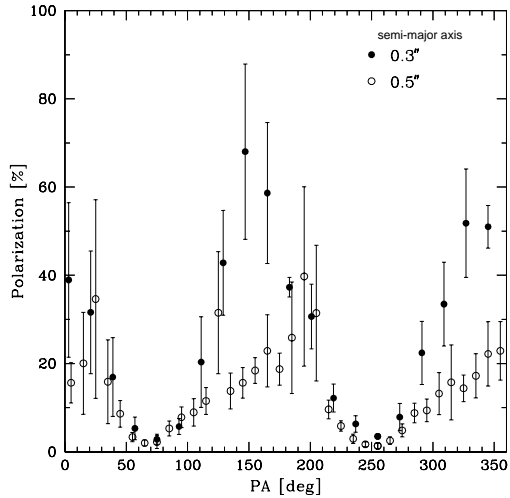}
  \end{center}
  \caption{
  	An azimuthal profile of the polarization degree of the circumstellar disk. 
        It is measured on the ellipse of eccentricity $e=0.72$ and a semi-major axis of 0.30$\arcsec$ and 0.50$\arcsec$, 
        which corresponds to the disk radius of  $\sim$ 40 AU and 70 AU. 
        For the position angle (PA), north is 0$\degree$and east is 90$\degree$. 
        The plots show mean polarization degrees binned in 9 $\times$ 9 pixels, and maxima and minima are clearly seen in 90$\degree$ intervals.
        }\label{fig:Pprof_obs}
\end{figure}

\begin{figure}
  \begin{center}
    \FigureFile(70mm,70mm){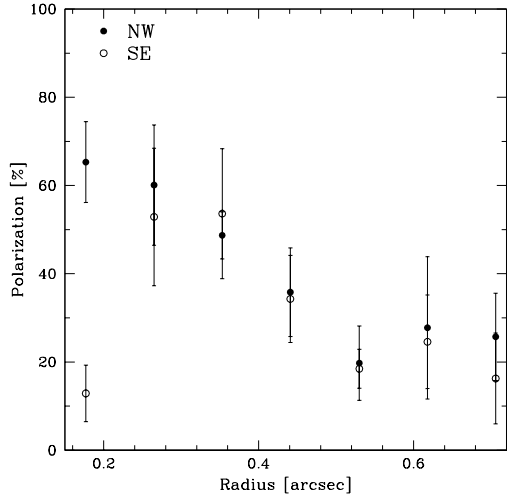}
  \end{center}
  \caption{
  	Radial profiles of the polarization degrees measured along semi-major axes of the circumstellar disk of UX Tau A. 
        The plots show mean polarization degrees binned in 9 $\times$ 9 pixels. 
        The filled and open circles denote the polarization degrees along semi-major axes in the north-west and south-east directions, respectively.
        From 0.15$\arcsec$ to 0.50$\arcsec$, the degrees of polarization show decreasing trends.
        }\label{fig:Pradprof}
\end{figure}

Figure \ref{fig:Pprof_obs} shows an azimuthal profile of the polarization degree ($P$) along the ellipse of $e = 0.72$ ($i = 46\degree$) 
and a semi-major axis of 0.30$\arcsec$ (42 AU). We measured mean polarization degrees and standard deviations in every 9 $\times$ 9 pixel region. 
It is sinusoidal; two maxima and two minima appear with 90$\degree$ intervals. This pattern is prominent up to the semi-major axis of $\sim$ 0.5$\arcsec$. 
Figure \ref{fig:Pradprof} represents radial profiles of the polarization degree along the semi-major axes. 
We measured mean polarization degrees and standard deviations in every 9 $\times$ 9 pixel region. 
The degree of polarization declines with radial distance to $\sim$ 0.5$\arcsec$. 
Beyond 0.8$\arcsec$ from the central star, the polarization is not significantly detected. 
% about PI & P error comments %
Assuming that the polarized intensity and the polarization degree of the disk are constant within binned 9 $\times$ 9 pixels respectively, 
these uncertainties are up to $\sim$ 80 ADU and $\sim$ 20 \% as shown in Figures \ref{fig:PIradprof} and \ref{fig:Pprof_obs}.
These estimates do not take into account any systematic effects from
the data reduction such as the PSF subtraction and the correction for
the instrumental polarization. Further discussion on the instrumentation
and the systematic errors will be presented elsewhere.

%% polarization by ISM 
Polarization due to interstellar medium in front of UX Tau binary system is negligible. 
The system is slightly embedded in a molecular cloud; the interstellar extinction of UX Tau A is 1.8 mag in the $V$-band (\cite{Esp10}). 
\citet{Serkowski75} obtained a relationship between polarization and reddening as $E(B-V) \geqq P/9$. 
We would therefore anticipate little foreground interstellar polarization would be present. 
Indeed, the polarization degree of UX Tau C, which was measured in 50 $\times$ 50 pixels  centered on the star, is only 0.5 \%.
We do not detect any circumstellar structures around UX Tau C. 
Since no 880 $\mu$m emission is detected from UX Tau C (\cite{And11}), this source is unlikely to have circumstellar structures 
and its measured polarization can be assumed to be a reasonable proxy for the interstellar polarization along the line of sight to UX Tau A.
Therefore, the interstellar medium in front of the UX Tau system has only negligible polarization. 

%% polarization by outflows??
Outflows emanating to the north-west and south-east directions are also possible sources to explain the elliptical structure. 
\citet{Lucas04} detected the polarization structure around HL Tau and suggested that 
its strong polarization structure is caused by bipolar outflows. 
However, the disk around UX Tau A has already been imaged by Sub-millimeter Array (\cite{And11}). 
The inclination and the position angle of the semi-major axis of the disk were determined to be 35$\degree$ and 176$\degree$, respectively. 
The disk geometry is roughly consistent with that derived from the polarized intensity image. 
Therefore, we consider that the elliptical structure we detected is not attributed to the outflow, but to the circumstellar disk around the central star.

\section{Discussion}
\subsection{Geometry of the circumstellar disk}
%% undetected gap and inner disk
The circumstellar disk looks smooth. 
A gap-like structure at 0.21 -- 30 AU is suggested from the disk model fitting to the SED (\authorcite{Esp10} \yearcite{Esp10}, \yearcite{Esp11}). 
A disk cavity is also resolved by the SMA observations (\cite{And11}), 
which is identical with the outer radius of the gap structure. 
However, we do not find such a structure. 
Our data do not show any signs for a gap or cavity down to the achieved inner working angle ($\sim$ 23 AU). 
We are not sensitive to a possible inner disk with the radius of $<$ 0.21 AU (\cite{Esp10}) because it is within the coronagraphic mask. 

%% UX Tau C effect to the disk
We examine whether UX Tau C influences the disk geometry of UX Tau A. 
The Lagrangian point L1 is the point where gravitational forces of the primary and tertiary stars 
balance each other to a mass point. With the primary mass of 1.3$M_{\odot}$ , the tertiary mass of 0.16$M_{\odot}$ (\cite{Kraus09}), 
and the separation of 2.63$\arcsec$ (\cite{White01}), the location of the point L1 was derived to be 1.9$\arcsec$ from the primary star. 
Given that the disk of UX Tau A extends up to 0.8$\arcsec$, we conclude that the disk is located 
in the gravitational zone of the primary star and is not gravitationally affected by UX Tau C.

\subsection{Polarization models of the circumstellar disk}
\begin{figure}
  \begin{center}
    \includegraphics[keepaspectratio]{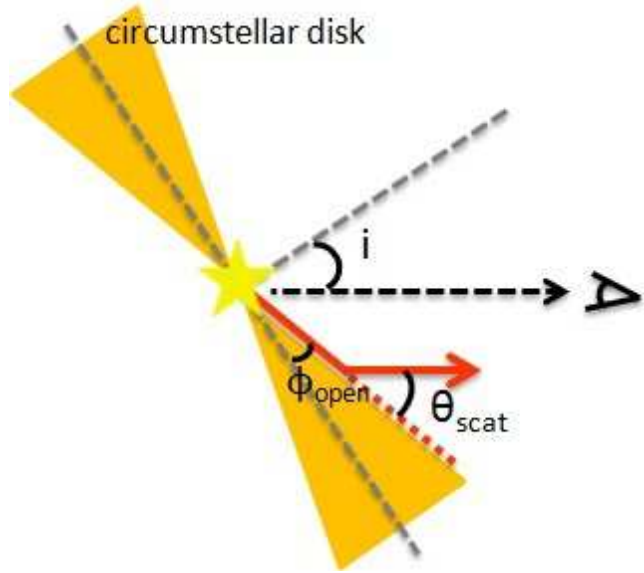}
  \end{center}
  \caption{Schematic view of the disk geometry.
  	}\label{fig:disk_geo}
\end{figure}

%% polarization model
We construct several polarization models of the circumstellar disk. First, we define the disk geometry. 
A scattering angle is an angle of deflection from the forward direction of incident light.
We consider the disk model based on \citet{McCabe02}.
In the disk geometry of Figure \ref{fig:disk_geo}, the scattering angle is calculated as
\begin{equation}
	\cos(\theta_{{\rm scat}}+\phi_{{\rm open}})=\sqrt{1-\frac{1}{1+\cos^2(PA-PA_{0})\tan^{2}i}}(-1)^{j}\label{eq:eqMc} ,
\end{equation}
where $i$ is the disk inclination to the line of sight, $PA$ is the position angle measured on the mid-plane of the disk, 
and $PA_{0}$ is the position angle on the near side of the semi-minor axis of the disk. 
$\phi_{{\rm open}}$ is the opening angle of the disk. It expresses the disk height with optical thickness $\tau >$ 1. 
If $\cos{PA} > 0$, $j=1$, and if $\cos{PA} < 0$, $j=0$ (\cite{McCabe02}).
Substituting the inclination, the semi-minor axis, and the height of the disk around UX Tau A into equation (\ref{eq:eqMc}) 
allowed us to acquire the scattering angle at each position angle. The disk inclination of 46$\degree$ was applied. 
From the semi-major axis of 165$\degree$ (Figure \ref{fig:PIimg}), we assumed that the semi-minor axis direction was 255$\degree$. 

%% Rayleigh scattering approach
Next, we consider three scattering approximations. 
The size parameter for dust grains is defined as $X = 2\pi a / \lambda$. 
$X = 1$ corresponds to 0.25 $\mu$m of the grain radius ($a$) for the $H$-band observations. 
We first adopt, as the simplest case, a Rayleigh scattering approximation. 
The Rayleigh scattering approximation holds in the case of sufficiently small dust grains compared to the observing wavelength ($X \ll 1$). 
For the $H$-band observations, it can be applied for the grains with the radii ($a$) much smaller than 0.25 $\mu$m. 
The relationship between the degree of polarization and the scattering angle is represented by the following equation (\cite{Hulst57}), 
\begin{equation}
	P=\frac{1-\cos^{2}\theta_{{\rm scat}}}{1+\cos^{2}\theta_{{\rm scat}}}\label{eq:eqRayleigh}\hspace{4pt}.
\end{equation}
The degree of polarization $P$ shows a maximum value $P_{{\rm max}}$ at $\theta_{{\rm scat}}$ = 90$\degree$ and 
a minimum value $P_{{\rm min}}$ at $\theta_{{\rm scat}}$ = 0$\degree$ and 180$\degree$. 
By combining equations (\ref{eq:eqMc}) and (\ref{eq:eqRayleigh}), we are able to calculate the polarization pattern on a circumstellar disk. 

\begin{figure}
  \begin{center}
    \FigureFile(70mm,70mm){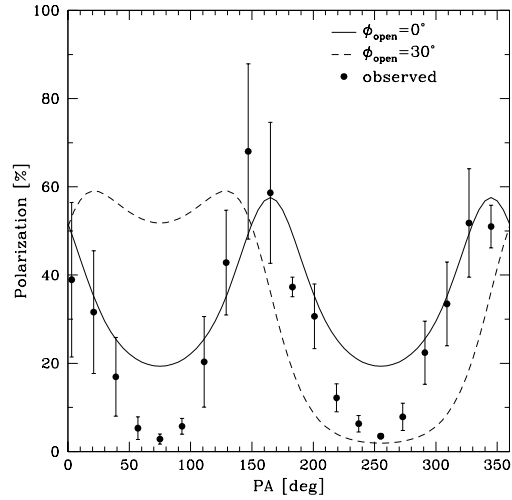}
  \end{center}
  \caption{
  	Azimuthal profiles of the polarization degree. The filled circles show the observed degrees of polarization 
        measured along the ellipse with an eccentricity of 0.72 and a semi-major axis of 0.30$\arcsec$. 
        The solid line represents the polarization degree of a geometrically thin disk model ($\phi_{{\rm open}} = 0\degree$) 
        and the dashed line shows that of a geometrically thick disk model of $\phi_{{\rm open}} = 30\degree$. 
        We used the Rayleigh scattering approximation. 
        The polarization profiles of the models are convolved with a 20$\degree$ window 
        in order to match the spatial resolution of the 9 $\times$ 9 pixels binned profile of the observational polarization.
        }\label{fig:Pmodl_R}
\end{figure}

We consider a geometrically thin disk ($\phi_{{\rm open}}$ = 0$\degree$). 
It produces a sinusoidal azimuthal profile of the polarization degree (Figure \ref{fig:Pmodl_R}). 
Maximum polarization degrees are $\sim$ 60 \% at $PA(P_{{\rm max}})$ $\sim$ 165$\degree$, 345$\degree$. 
Minimum degrees are $\sim$ 20\% at  $PA(P_{{\rm min}})$ $\sim$ 75$\degree$, 255$\degree$. 
Two $P_{{\rm max}}$ and two $P_{{\rm min}}$ have the same values respectively. 
The interval between $PA(P_{{\rm max}})$ and $PA(P_{{\rm min}})$ is 90$\degree$. 
Although the data points match the observational polarization degrees quite well for $P_{{\rm max}}$, 
they shows large differences (about 20 \%) from the observational degrees  for $P_{{\rm min}}$. 
The case of a geometrically thick disk with $\phi_{{\rm open}} = 30\degree$ is also shown in Figure \ref{fig:Pmodl_R}. 
Whereas two $P_{{\rm max}}$ have the same values, two $P_{{\rm min}}$ have different values. 
Thus, we conclude that neither disk model matches the observational azimuthal profile of the polarization degrees. 
%% grain sizes Espaillat estimated
\citet{Esp11} suggested that maximum grain size is 10 $\mu$m at the inner edge of the inner disk 
and 5 $\mu$m at the inner edge of the outer disk, from fitting a disk model to the SED of UX Tau A. 
If the disk consists only of such large dust grains, the Rayleigh scattering approximation is not appropriate for the $H$-band observations. 

\begin{figure}
  \begin{center}
    \FigureFile(70mm,70mm){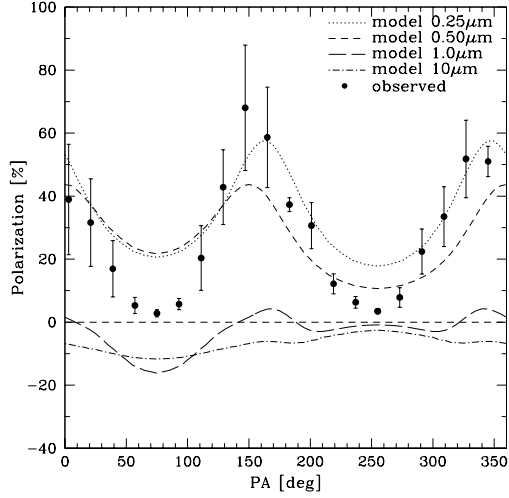}
  \end{center}
  \caption{
  	Azimuthal profiles of the polarization degree. The filled circles show the observed degrees of polarization 
        measured along the ellipse with an eccentricity of 0.72 and a semi-major axis of 0.30$\arcsec$. 
        We constructed four geometrically thin disk models with the maximum grain radius of 0.25, 0.50, 1.0, and 10 $\mu$m, using the Mie scattering approximation.
        The polarization profile of each model is shown by the dot, the short dashed, the long dashed, the dashed dotted lines, respectively. 
        We used 0.005 $\mu$m for the minimum grain radius and $N(a) \propto a^{-3.5}$ for the grain size distribution.
        The polarization profiles of the models are convolved with a 20$\degree$ window 
        in order to match the spatial resolution of the 9 $\times$ 9 pixels binned profile of the observational polarization.
        }\label{fig:Pmodl_M}
\end{figure}

%% Mie scattering approach
Next, we consider a Mie scattering approximation, which is applicable for spherical dust grains 
with the size comparable to the observing wavelength ($X \approx 1$, i.e. $a$ $\sim$ 0.25 $\mu$m). 
The degree of polarization depends not only on the scattering angle, 
but also on the grain size, the size distribution, the refraction index, and other grain parameters.
Using the observing wavelength, the grain size, the grain size distribution, the refraction index ($n$), and the absorbing coefficient ($k$), 
the Mie scattering program allows us to obtain the degree of polarization as a function of the scattering angle (Figure \ref{fig:Pmodl_M}). 
A similar work done on comet theory using Mie theory, details the procedure (\cite{Sen91}). 
We found that large dust grains show low $P_{{\rm max}}$. 
In the case that the largest dust grains have a radius as large as 1 mm, 
the polarization degree shows negative for the large scattering angle (\cite{Murakawa10}). 
The negative polarization indicates a radial pattern of the polarization vectors. 
We conclude that any Mie scattering approximation does not reproduce the observed polarizations of UX Tau A. 
Even if we use a geometrically thick disk, the models are hardly different from that of a geometrically thin disk. 

\begin{figure}
  \begin{center}
    \FigureFile(70mm,70mm){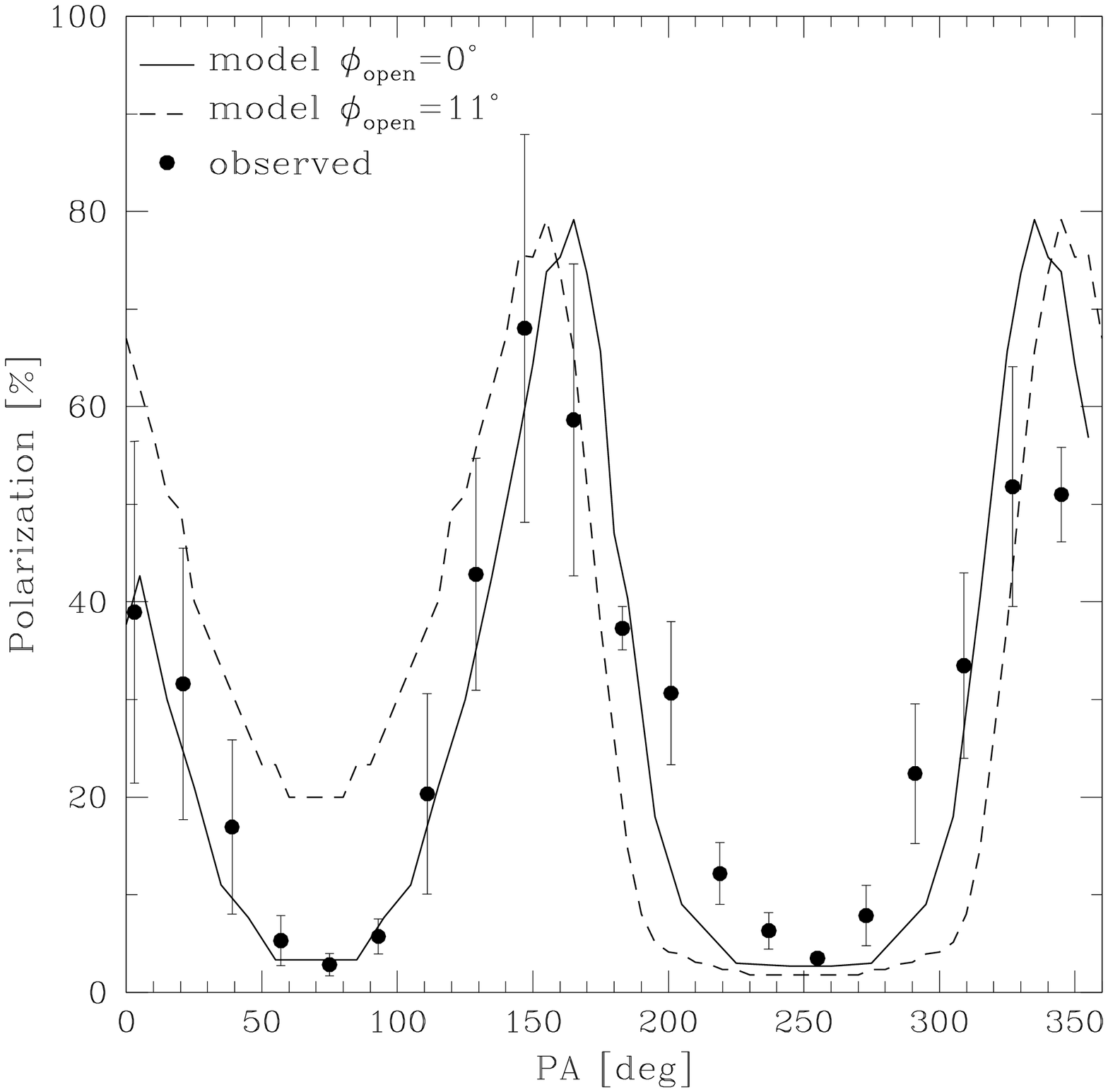}
  \end{center}
  \caption{
  	Azimuthal profiles of the polarization degrees of the circumstellar disk. 
        The filled circles show the observed degrees of polarization 
        measured along the ellipse with an eccentricity of 0.72 and a semi-major axis of 0.30$\arcsec$. 
        We showed two disk models constructed by the geometric optics approximation.
        The solid line represents the geometrically thin disk model. 
        The dashed line shows the geometrically thick disk model with $\phi_{{\rm open}}$ = 11$\degree$. 
        The polarization profiles of the models are convolved with a 20 $\degree$ window. 
        Both models approximately reproduce the observed polarization degrees.
	}\label{fig:Pmodel_GO}
\end{figure}

%% Geometric optics approach
Finally, we focused on geometric optics. It is applicable for interpreting light scattering 
by irregular shaped particles sufficiently larger than the observing wavelength ($X \gg 1$, i.e. $a\gg$ 0.25 $\mu$m). 
We adopted computer simulation results of \citet{Grynko03}, 
in which light scattering is calculated for various grains. 
We used the polarization profiles for 100 faceted particles. 
\citet{Grynko03} calculated it with the refractive indices of some silicate ($n=1.5$, $k=0.004$), and $a\sim$ 30 $\mu$m. 
We considered that the 100 faceted particles presumably correspond not to the dust grains with smooth spheres but to those with rugged surfaces. 
The degree of polarization shows 0 \% at $\theta_{{\rm scat}}$ = 0$\degree$, 180$\degree$ and maximum value at $\theta_{{\rm scat}}$ $\sim$ 85$\degree$. 
These distribution of the polarization degree resembles that of the Rayleigh scattering approximation. 
One characteristic is that the degree of polarization falls down to about 10 \% at $\theta_{{\rm scat}}$ $\sim$ 45$\degree$ and 135$\degree$,
whereas that in the Rayleigh scattering approximation has about 30 \% at $\theta_{{\rm scat}}$ = 45$\degree$ and 135$\degree$.
With this distribution of the polarization degrees, we constructed the polarization model of a geometrically thin disk (Figure \ref{fig:Pmodel_GO}). 
Maximum polarization degrees are $\sim$ 80\% at $PA(P_{{\rm max}})$ = 165$\degree$ and 335$\degree$. 
Minimum degrees are $\sim$ 3 \% at $PA(P_{{\rm min}})$ = 55$\degree$ -- 85$\degree$ and 225$\degree$ -- 275$\degree$. 
Although the model shows maximum degrees about 10 \% larger than the observed maximum degree at $PA$ $\sim$ 165$\degree$, 
it reproduces well the observed minimum polarization degree. 
Moreover, $PA(P_{{\rm max}})$ and $PA(P_{{\rm min}})$ of the model are consistent with those of the observed $PA$s. 
We did not obtain such a profile in the Mie scattering approximation by spherical grains with $a$ = 30 $\mu$m. 
Therefore, we suggest that dust particles in the disk are nonspherical grain with 60 $\mu$m diameter at 40 AU from the central star. 
This argument is consistent with no 10 $\mu$m silicate emission in the UX Tau A spectra, indicating a lack of small dust grains (\cite{Esp10}).

We also considered a geometrically thick disk. Assuming vertical hydrostatic equilibrium and no turbulence in the disk, 
the density profile of the disk is given by
\begin{equation}
	\rho(Z)=\rho_{0}\exp{\left(-\frac{Z^{2}}{H^{2}}\right)}\label{eq:eqdensity} ,
\end{equation}
where $H$ is the scale height and $\rho_{0}$ is the density at the disk mid-plane. 
\citet{Muto11} indicates that the disk thickness $Z$ with $\tau = 1$ corresponds to 
$3H$ for such a disk. 
The large opening angle of the circumstellar disk, $\phi_{{\rm open}}$, 
corresponds to large ratio of the scale height to the disk radius ($H/R$). 
We simulated thick disks with various $\phi_{{\rm open}}$ values. 
A disk with large $\phi_{{\rm open}}$ value shows large interval of $PA(P_{{\rm max}})$. 
For example for the disk with $H/R$ of 0.1, $PA(P_{{\rm max}})$ results in 150$\degree$ and 350$\degree$.
Moreover, minimum polarization degrees at $PA(P_{{\rm min}})$ = 70$\degree$ and 250$\degree$ have different values. 
It is distinctly different from the polarization profile of UX Tau A. If $\phi_{{\rm open}} <$ 11$\degree$, 
corresponding to $H/R <$ 0.067, the models are consistent with the polarization profile of UX Tau A. 
We conclude that the geometrical thickness of the UX Tau A disk agrees with $H/R$ = 0 -- 0.067. 
Most of the dust grains in the disk have therefore settled into the mid-plane and formed the flat disk. 
\citet{And11} found a disk geometry favoring a rather flat disk structure. 

The polarization degrees decrease with the semi-major axes (Figure \ref{fig:Pradprof}).
Since the Keplerian velocity is small in the outer part of the disk, 
it is generally considered that the dust particles in the outer disk have small relative velocities. 
We consider a slow growth rate of such dust, thus resulting in small dust grains in the outer part of the disk. 
However, the low polarization degrees in the outer part of the disk are not reproduced by the Rayleigh scattering approximation. 
The computer simulations of light scattering with geometric optics approximation (\cite{Grynko03}) 
calculated the polarization degree profile for grains with various number of facets. 
They indicated that the maximum polarization degree decreases with decreasing the facet number. 
We expect dust grains with rough surface in the outer part of the circumstellar disk. 
% plus arguments 
Most previous studies of scattered light images from circumstellar disks could explain 
the scattering (or polarization) function with small, ISM-like grains. 
For AB Aur, for instance, \citet{Perrin09} found a similar $P_{\rm max}$ as in this present study, 
but they could explain the results with more or less typical, small ISM grains. 
Only recently, \citet{Quanz11} found some indications based on PDI data, 
that on the surface of the HD100546 disk micron sized grains might be present. 
\citet{And11} used also a rather ISM-like dust population to explain both their interferometric observations and the SED of UX Tau A. 
However very small $P_{\rm min}$ observed in this work can be reproduced by the scattering model of the dust grains with rough surfaces. 
To obtain further information of shapes and sizes of the grains, geometric optics models are to be investigated with various parameters.
% from here

\subsection{Dust growth in the circumstellar disk}
%% Theoretical discussion
%% dust sizes at the age of UX Tau A
The forces acting on dust grains in a protoplanetary disk are gas drag force and gravitational force of the central star. 
In a condition with a temperature typically expected in a circumstellar disk, grains with the sizes less than a few tens micrometers are marginally affected 
by gas drag forces and dominated by Brownian thermal motion (\cite{Weiden00}). 
As a consequence of frequent collision and coagulation, the dust grains grow larger. 
We examine how large dusts can grow during settling into the mid-plane at 40 AU from the central star. 
It is claimed that the dust grains grow to several centimeters at 1 AU, during sedimentation (\cite{Nakagawa81}). 
We calculated the radii of grains settling toward the mid-plane according to the equations for dust growth during sedimentation (\cite{Takeuchi09}). 
We assumed the dust sticking probability $C = 1$ and the grain bulk density of 1 g cm$^{-3}$. 
We do not consider any turbulence in the disk. As a result, we found that the dust grains can grow up to 100 $\mu$m radius during sedimentation at 40 AU from the central star. 
Therefore, it is possible for dust grains to grow to 30 $\mu$m radius at 40 AU from the central star.

%% dust growth timescale
Next we evaluate the timescale that the dust grains grow from 0.1 $\mu$m to 100 $\mu$m 
during sedimentation toward the disk plane. We take account of the gas drag. 
If the circumstellar disk of UX Tau A is optically thin at sub-millimeter wavelengths, 
the sub-millimeter observations (\cite{And11}) indicate that 
the gas surface density of the UX Tau A disk at 40 AU from the central star is $\sim$ 10 g cm$^{-2}$. 
For the disk with $H/R =$ 0.067, the disk thickness with $\tau =$ 1 is about 16 AU at the disk radius of 40 AU. 
We obtain the average gas density of 4.2 $\times$ 10$^{-14}$ g cm$^{-3}$ (2.5 $\times$ 10$^{10}$ atom cm$^{-3}$). 
With the relationship between the gas density and the mean free path of gas molecules (\cite{Takeuchi09}), 
we derive 4.8 $\times$ 10$^{4}$ cm for the gas mean free path. 
Since the radii of the dust grains up to 100 $\mu$m is sufficiently smaller than the gas mean free path, 
the dust grains grow under the Epstein regime. 
As a consequence of calculating the equation for the sedimentation timescale (\cite{Takeuchi09}), 
it is revealed that it takes at least 10$^{5}$ years for dust grains with 0.1 $\mu$m radii to settle and grow up to 100 $\mu$m in radius. 
The age of UX Tau A is estimated to be about 10$^{6}$ years (\cite{Kraus09}). 
We claim that the coagulated growing dusts have already settled into the equatorial plane of the circumstellar disk. 
It is consistent with our observational evidence that the circumstellar disk of UX Tau A is geometrically thin.

%% radial fall timescale
Dust grains with sizes between a few micrometer and a hundred micrometer are entrained by gas toward the central star. 
\citet{Adachi76} investigated the spiral motion of the dust grains in a protoplanetary disk. 
The decay time of the spiral motion corresponds to the infall timescale of the dust grains toward the central star. 
Applying their calculations, 
with the radially constant gas density of 4.2 $\times$ 10$^{-14}$ g cm$^{-3}$, 
we derived 2 $\times$ 10$^{6}$ years for the dust grains with radii of 100 $\mu$m to infall from 40 AU to the central star.
We therefore suggest that the circumstellar disk of UX Tau A still contains such large dust grains.

\section{Conclusions}

With HiCIAO/AO188 mounted on the Subaru Telescope, we carried out $H$-band polarimetric 
imaging observations of UX Tau A, which has been classified as a pre-transitional disk object. 
The observation revealed a circumstellar disk around UX Tau A at the spatial resolution of 0.1$\arcsec$ beyond 23 AU from the central star. 
The disk extends to 120 AU in radius. The disk is inclined by 46$\degree$ $\pm$ 2$\degree$ as the west side is near. 
We did not detect evidences of the inner disk and the gap-like structure. 

The notable feature of the circumstellar disk is the huge variation of the polarization degrees. 
It varies from 1.6 to 66 \%. 
We constructed several polarization models of the circumstellar disk 
based on the Rayleigh scattering and Mie scattering approximations. 
However neither models were consistent with the observational azimuthal profile of the polarization degrees. 
Focusing on geometric optics, we built the polarization model of the geometrically thin disk with nonspherical grains with the radii of 30 $\mu$m. 
The model reproduced well the observational azimuthal profile of the polarization degree. 
We suggest that UX Tau A has a geometrically thin disk containing the nonspherical dust grains with the radii of 30 $\mu$m. 

Such a disk with nonspherical large dust grains is consistent with the core accretion model of planetary formation process. 
At 40 AU from UX Tau A, the dust grains can grow up to 100 $\mu$m 
in radius by collisional coagulation and settle toward the mid-plane with the timescale of 10$^{5}$ years at the earliest. 
Observational evidence of large dust grains as well as the gap structure in the circumstellar disk 
provides robust signatures of planetary formation process in the UX Tau A system.

%%%%%%%%%%%%%%%%%%%%%%%%%%%%%%%%%%%%%%%

%\begin{longtable}{lll}
%  \caption{Sample of ``longtable"}\label{tab:LTsample}
%  \hline              
%  name & value1 & value2 \\ 
%\endfirsthead
%  \hline
%  name & value & value2  \\
%\endhead
%  \hline
%\endfoot
%  \hline
%\endlastfoot
%  \hline
%  aaaaa & bbbbb & ccccc \\
%  ...... & ..... & ..... \\
%  ...... & ..... & ..... \\
%  ...... & ..... & ..... \\
%  xxxxx & yyyyy & zzzzz \\
%\end{longtable}

\bigskip

%Acknowledgement should be placed at end of main text.
%(NOT after the Appendix.)
We thank the telescope staff members and operators at the Subaru Telescope.
This work is partly supported by the JSPS-DST collaboration.
E.L.T. gratefully acknowledges support from a Princeton University Global Collaborative Research Fund grant 
and the World Premier International Research Center Initiative (WPI Initiative), MEXT, Japan. 
J. Carson gratefully acknowledges support from NSF grant AST-1009203.

%\appendix
%\section{Method of .....}

%\section{Approximation of ...}

%\section*{Complete data}

%%%
% See the manual for the detail.
%%%

\end{document}